\documentclass[aps,prr,twocolumn,groupedaddress,superscriptaddress]{revtex4-2}
\usepackage{graphicx}
\usepackage{amssymb,amsmath}
\usepackage{xcolor}
\usepackage{bm}
\usepackage{epstopdf}
\usepackage{url}

\begin{document}

\title{Efficient shortcuts-to-adiabaticity for loading an ultracold Fermi gas into higher orbital bands of one-dimensional optical lattice}

\author{Hang Yu}
\thanks{These authors contributed equally to this work.}
\affiliation{School of Physics and Astronomy, Sun Yat-Sen University, Zhuhai 519082, China}
\author{Haoyi Zhang}
\thanks{These authors contributed equally to this work.}
\affiliation{School of Physics and Astronomy, Sun Yat-Sen University, Zhuhai 519082, China}
\author{Bolong Jiao}
\affiliation{School of Physics and Astronomy, Sun Yat-Sen University, Zhuhai 519082, China}
\author{Qinxuan Peng}
\affiliation{School of Physics and Astronomy, Sun Yat-Sen University, Zhuhai 519082, China}
\author{Liao Sun}
\affiliation{School of Physics and Astronomy, Sun Yat-Sen University, Zhuhai 519082, China}
\author{Jiaming Li}
\email[]{lijiam29@mail.sysu.edu.cn} 
\affiliation{School of Physics and Astronomy, Sun Yat-Sen
University, Zhuhai 519082, China} 
\affiliation{Shenzhen Research Institute of Sun Yat-Sen University, Shenzhen 518087, China} 
\affiliation{Guangdong Provincial Key Laboratory of Quantum Metrology and Sensing, Sun Yat-Sen University, Zhuhai 519082, China}
\affiliation{State Key Laboratory of Optoelectronic Materials and Technologies, Sun Yat-Sen University, Guangzhou 510275, China} 
\author{Le Luo}
\email[]{luole5@mail.sysu.edu.cn} 
\affiliation{School of Physics and Astronomy, Sun Yat-Sen
	University, Zhuhai 519082, China} 
\affiliation{Shenzhen Research Institute of Sun Yat-Sen University, Shenzhen 518087, China} 
\affiliation{Guangdong Provincial Key Laboratory of Quantum Metrology and Sensing, Sun Yat-Sen University, Zhuhai 519082, China}
\affiliation{State Key Laboratory of Optoelectronic Materials and Technologies, Sun Yat-Sen University, Guangzhou 510275, China} \affiliation{Quantum Science Center of Guangdong-HongKong-Macao Greater Bay Area, Shenzhen 518048, China}

\date{\today}

\begin{abstract}

We propose an experimental scheme to load ultracold Fermi gases from the ground orbital band of a one-dimensional optical lattice into the first excited orbital band. Unlike the narrow momentum distribution of a Bose-Einstein Condensate, Fermi gases exhibit a broad momentum distribution. To address this, we define the average loading efficiency across all quasi-momentum states and theoretically perform the loading operation simultaneously for each Bloch state. Using a multi-parameter global optimization method, we determine the loading efficiency at various lattice depths. We can enhance the loading efficiency by adjusting the phase of the lattice, which leverages the different symmetries of Bloch wavefunctions in various optical lattice orbitals. We also identified that the primary factor hindering higher loading efficiency in the Fermi gas is the multiple occupancy of the quasi-momentum states. Our simulations of various occupancies revealed a decreasing trend in mean loading efficiency as the number of occupied quasi-momentum states increases. Finally, we compare our method with other loading techniques and assess its experimental feasibility.

\end{abstract}

\maketitle


\section{Introduction}
Exploring quantum phenomena in solid-state systems remains a long-standing and crucial goal in modern physics~\cite{Bloch2008,Sakurai2017}. Understanding the complex interactions in materials with multiple degrees of freedom, such as spin, charge, and orbital, is vital for advancing both fundamental research and quantum technologies.
One of the most promising platforms for studying these interactions is the cold atoms in an optical lattice, which offers unparalleled control over atomic interactions, lattice structures, and external force fields~\cite{Bloch2005,Lewenstein2007,Dutta2015,Gross2017,Schafer2020}. This precise control enables the simulation of quantum systems that are difficult to investigate through traditional condensed matter experiments~\cite{Li2016}. 

Historically, research on cold atom optical lattices has primarily focused on atoms in the $s$-band of optical lattices due to experimental limitations~\cite{Gunter2005,Stoferle2006,Had06,Martiyanov2010,Frohlich2011,Feld2011,Boettcher2016,Waseem2016,Waseem2017,Holten2018,Waseem2019,Murthy2019,Chang2020,Marcum2020}.
However, recent advances have allowed the exploration of atoms in higher bands, such as $p$- and $d$-bands, opening new avenues for studying quantum phenomena like orbital physics~\cite{Wirth2011,Jin2021,Wang2023,Hu2018,Hachmann2021}, unconventional superfluidity~\cite{Olschlager2011,Niu2018,Wang2021}, complex interaction dynamics~\cite{Venu2023,Jackson2023,Kiefer2023} and superconductivity~\cite{Dong2024,Luo2023,Zhang2023,Nakayama2021,Hu2015PRX}.

Several methods, including stimulated Raman transitions~\cite{Muller2007,Venu2023}, population exchange technique~\cite{Hachmann2021,Wang2021,Kiefer2023}, and modulation of optical dipole trap (ODT) depth~\cite{Gong2023,Jackson2023,Dale2024}, as well as various additional approaches~\cite{Denschlag2002,Taie2015,Kohl2005}, have been developed to load cold atoms into higher bands. In the stimulated Raman transition, laser fields are used to transfer atoms from the ground band to excited bands, which allows for precise control without significantly heating the system. In Ref.~\cite{Muller2007}, this method has successfully been used to load atoms into the $p$-band, leading to the observation of high-orbital Bose-Einstein Condensate (BEC). In population exchange technique, atoms are often transferred under external energy levels modulation to higher bands through interaction-driven processes. In Ref.~\cite{Kiefer2023}, this technique is implemented to study molecule formation in higher orbitals, revealing new forms of interaction in multi-orbital systems. By adjusting the depth of the ODT, researchers can also change the chemical potential of the system, facilitating the occupation of higher bands. In Ref.~\cite{Gong2023}, this method demonstrates controlled loading into higher bands, providing a tunable platform for studying high-orbital dynamics.

While the above techniques have been effective for boson gases, efficiently loading Fermi gases into high bands remains challenging due to differences in wave function symmetry and the nonzero momentum distribution of Fermi gases in optical lattices. For example, in the $s$-band, Fermi gases exhibit broad momentum distributions, with atoms occupying all quasi-momentum states, which complicates the efficient loading with a single technique. Therefore, developing an easy-to-use and efficient method to load Fermi gases into high bands is urgently needed. Such a technique would significantly advance studies of anisotropic interactions~\cite{Venu2023,Jackson2023}, orbital ordering, and the realization of exotic quantum phases.

\begin{figure}[htbp]
	\begin{center}
		\includegraphics[width=\columnwidth, angle=0, scale=1.0]{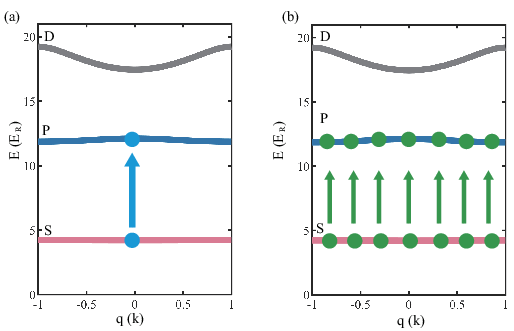}
		\caption{
			Loading a BEC and a Fermi gas into an optical lattice differs significantly, as shown schematically with band diagrams:
			(a) For a BEC, the $s$-band momentum distribution is narrow, centered at zero quasi-momentum, with only a single state transferred to the $p$-band.(b) For Fermi gases, the $s$-band momentum distribution is broad, and all quasi-momentum states are independently loaded into the $p$-band without interference.
		}    \label{p:Schematic Diagram}
	\end{center}
\end{figure}

Building on our prior work~\cite{Gong2023} on producing variable band ratios to study the crossover from two dimensions to three dimensions (2D-3D crossover) dynamics in optical lattices, this work presents a method to load ultracold Fermi gases into the first excited state ($p$-band) of a one-dimensional optical lattice. Inspired by the shortcuts-to-adiabaticity method developed in Ref.~\cite{Liu2011,Zhou2018,Hu2015,Yang2016,Wang2016,Guo2021,Odelin2019,Jin2022}, our approach involves four key steps: (1) transferring atoms from the ODT to the optical lattice $s$-band; (2) modulating the optical lattice laser intensity to induce energy variations and modify atomic Bloch states; (3) shifting the optical lattice potential to adjust the phase of the Bloch states to match the $p$-band parity; and (4) continuously adjusting the lattice laser to achieve optimal alignment of the Bloch states with the $p$-band. Given the broad momentum distribution of Fermi gases~\cite{Modugno2003}, we adopt a strategy to simultaneously load atoms across all quasi-momentum states and perform global optimization throughout the process to achieve the optimal time sequence, as shown in Fig.~\ref{p:Schematic Diagram}. With our method, the highest loading ratio for $p$-band reaches $95\%$, improving upon efficiencies reported in earlier works. It offers reduced loading times and simpler implementation compared to existing techniques. Our method is significant for extending the lifetime of high-orbital atoms and advancing research in high-orbital physics.

The structure of this paper is as follows: In Section~\ref{sec:theory}, we present the theoretical model of our optimization scheme. Section~\ref{sec:simulation} presents the details of the multi-parameter optimization method, including optimization using only optical pulse durations, optimization with pulse durations and a single lattice potential phase, and optimization with pulse durations alongside a phase tailored to each pulse. In Section~\ref{sec:discussion}, we discuss the effects of quasi-momentum occupancy on the result of the optimization, and compare our results with other higher band loading methods. Finally, we summarize the key findings of this work and offer perspectives for future research in Section~\ref{sec:summary}.

\section{Model}
\label{sec:theory}

In this section, we outline the theoretical framework underlying our loading method. To analyze the Hamiltonian, Bloch states, and time evolution operator, we employ the plane wave decomposition method, a widely adopted approach in the field. A detailed explanation of this method can be found in Appendix \ref{app:method}.
 
As a starting point, we first describe the initial state of the Fermi gas system. Building on our previous work, we have shown that nearly $100\%$ of the atoms can be loaded into the $s$-band by adiabatically raising the optical lattice~\cite{Gong2023}. Unlike a BEC, which occupies a single momentum state, the Fermi gas exhibits a broad momentum distribution, as shown in  Fig.~\ref{p:Schematic Diagram}. As a result, the initial state of the loading process can be expressed as follows:
\begin{equation}
{\psi _{in}} = \sum\limits_q {{C_{in}}\left| {s,q} \right\rangle }, 
\label{eq:4}
\end{equation}
$C_{in}$ representing the equal probability of $s$-band Bloch states for each $q$, $s$ means the $s$-band. 

Similarly, we express the target state we aim to achieve:
\begin{equation}
	{\psi _{t}} = \sum\limits_q {{C_{t}}\left| {p,q} \right\rangle }, 
	\label{eq:5}
\end{equation}
here, we assume that Bloch states in the $p$-band with an equal probability for each quasi-momentum, and this probability is identical to the one associated with the $s$-band mentioned above, i.e., $C_{t} = C_{in}$.

Thus, during the optical lattice switching process, each quasi-momentum state participates in the temporal evolution, which can be expressed as:
\begin{equation}
{\psi _f} = \prod\limits_{j = n_t}^1 {{{\hat U}_j}} {\psi _{in}},
\label{eq:6}
\end{equation}
where, $\hat{U}_j=e^{-i \hat{H}_j t_j}$ is the time evolution operator for the j-th process, where $n_t$ is the total number of processes. 

Based on the different time sequences, we divide the Hamiltonian into two scenarios: the shutdown and opening of the lattice potential, here are the Hamiltonians: 
\begin{equation}
    {H_s} = \frac{{{p^2}}}{{2m}}, {H_o} = \frac{{{p^2}}}{{2m}} + {V_o}(x),
	\label{eq:Hamiltonian}
\end{equation}
where ${H_s}$ corresponds to the lattice potential being off, and ${H_o}$ describes the lattice being on with a constant depth throughout the process. The final state is obtained after time evolution. 

After the evolution, we calculate the fidelity between the evolved final states and the target states for different $q$ values, and then obtain the average fidelity:
\begin{equation}
	F = \left|\left\langle\psi_t \mid \psi_f\right\rangle\right|^2.
	\label{eq:7}
\end{equation}
This average fidelity quantifies the proportion of atoms loaded into the $p$-band, which we aim to maximize through optimization. Using the \texttt{MATLAB} global optimization package, we define variables as time points and optimize the output to achieve the highest fidelity.

To facilitate efficient loading into $p$-band, it is necessary to adjust the parity of the atoms due to the parity difference between bands. This can be achieved through optical lattice phase modulation, as the parity of Bloch states at $q=0$ is even in the $s$-band and odd in the $p$-band~\cite{Zhou2018}. Zhou demonstrated that a phase modulation of $3\pi/4$ effectively achieves this adjustment, making it essential for successful $p$-band loading.

As noted earlier, the Fermi gas occupies nearly all quasi-momentum states, most of which lack well-defined parity except at $q = 0$. Moreover, the exact phases of the occupied quasi-momentum states are unknown due to their broad distribution. To address this challenge, we incorporate a phase $\varphi$ into the optimization program, where we add $\varphi$ to the modulated lattice potential $V^{\prime}(x)$ compare to the original lattice potential $V_o(x)$,
\begin{equation}
	V^{\prime}(x) = -s E_R \cos ^2\left(\frac{\pi x}{d}+\varphi \right),
	\label{eq:8}
\end{equation}
and the Hamitonian is:
\begin{equation}
	{H_m} = \frac{{{p^2}}}{{2m}} + V^{\prime}(x).
	\label{eq:phase change H}
\end{equation} 
Fig.~\ref{p:result}(a) shows the time sequence of the scheme, which is divided into five constrained segments. This segmentation represents a balance between achieving high loading efficiency and minimizing optimization time. The five segments correspond to different stages of switching the lattice potential on and off, with phase changes occurring at three yellow points to enhance loading efficiency.

\section{Method and Results}
\label{sec:simulation}

In this section, based on the experimental setup outlined in Appendix~\ref{app:experiment setup}, we present the method and results of loading the fermionic atoms into higher orbital band. We model the entire optimization process as a black-box function, with the initial state occupying all quasi-momentum states of $s$-band. Then we use the time evolution operator in Section~\ref{sec:theory} to describe the time evolution. Finally, the fidelity, representing the loading efficiency, is calculated according to the time sequences. To obtain the maximum fidelity we optimize the time sequences where the \texttt{Global Optimization Toolbox} in \texttt{MATLAB}~\cite{MATLAB2024} is used.

We prioritize time constraints due to the large adjustment range for lattice depth and phase shift (from $0$ to $\pi$). Since the lattice must be turned off during experiments, atoms undergo free ballistic expansion, so we limit the off-time to prevent atoms from escaping. Using the ballistic expansion formula:  	$b_z(t)=b_{0 z} \sqrt{1+\left(\omega_z t\right)^2}$ and considering a lattice size of $87 ~\mathrm{\mu m}$, we determine that the maximum off-time for a lattice depth of $15 E_R$ is $400 ~\mathrm{\mu s}$, with the maximum optimization time fixed at $226 ~\mathrm{\mu s}$. In the following, we conduct global optimization in three scenarios: the first with $\varphi = 0$ for all five time segments, the second with single-phase optimization, and the third with multi-phase optimization.

\begin{figure}[htbp]
	\begin{center}
		\includegraphics[width=\columnwidth, angle=0]{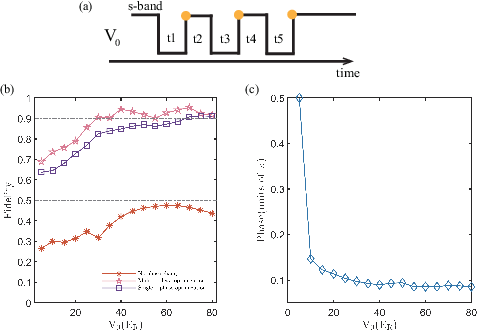}
		\caption{The results of the optimization of loading efficiency. (a) The yellow dots represent the time points at which we optimized the phase. In the single-phase optimization, the three yellow dots correspond to the same phase, while in the multi-phase optimization, each of the yellow dots represents a different phase. (b) The orange line represents the first scenario, with a maximum fidelity of approximately $48\%$. The purple line represents single-phase optimization, and the pink line represents multi-phase optimization, with the latter reaching a peak fidelity of up to $95\%$. The vertical axis value of $1$ corresponds to $100\%$, and this scale applies to all other cases as well. (c) For single-phase optimization, the optimized phase exhibits a monotonic decrease as a function of the lattice depth.}     
		\label{p:result}
	\end{center}
\end{figure}

In the first scenario, the phase of all the five segments remains zero degree. We observe the loading efficiency increases from $20\%$ to nearly $48\%$ where the lattice depth is rased from 5$E_R$ to 60$E_R$. The loading efficiency slightly decreases at higher depths (Fig.~\ref{p:result}(b), orange line). However, even at its best in this scenario, the efficiency falls short of the best in the BEC case at Ref.~\cite{Zhou2018}, motivating us to explore phase optimization in the next scenario.

The second scenario is the single-phase optimization, where we optimize a fixed phase $\varphi$ that remains consistent across all five time segments for the lattice potential. The results are shown in Fig.~\ref{p:result}(b) (purple line), which demonstrate an overall enhancement in the loading efficiency, reaching up to $90\%$ at the optimum. At lattice depths larger than $40 E_R$, the increase of the loading efficiency slows down, showing a saturation effect. This behavior can be attributed to the fact that as the lattice depth increases, the differences in the plane-wave distribution among the different Bloch states within the same band diminish. As shown in Fig.~\ref{p:result}(c), for different lattice depths, the optimized phase varies. At the shallowest lattice depth of $5 E_R$, the optimized phase is approximately $\pi/2$, and decreses to about $\pi/10$ with the increase of the lattice depth up to $30 E_R$.  For comparison, we show the result of fixing the phase to $3 \pi /4$ in Appendix~\ref{app:3pi/4}, which phase had been used to optimize the $p$-band loading of BEC in Ref.~\cite{Zhou2018}. With this fixed phase, the maximum loading efficiency of the Femi gas is approximately $55\%$, significantly lower than the results presented here. This highlights the importance of including the phase as an optimization parameter to enhance the loading efficiency of the Fermi gas. The phase optimization is more important for the shallow lattice depths, since the distribution of  quasi-momentum states are wider than the case of the deep lattice case. 

To achieve higher loading efficiency, we implemet the third scenarion of optimizing the phase at each instance when the lattice potential is turned on. This multi-phase optimization yields a maximum efficiency of nearly $95\%$, as shown by the pink line in Fig.~\ref{p:result}(b), and outperforms the second scenario. The efficiency generally increases with the lattice depth, reflecting the better performance under the  deeper potentials.  It is noted  this result represents the best efficiency achievable within the constraints of our system, rather than the absolute theoretical limit. The high-dimensional parameter space and the inherent complexity of the black-box function make achieving the true global optimum challenging. Nevertheless, this result is significant for achieving a high-purity population of atoms in the $p$-band state.

\begin{figure}[htbp]
	\begin{center}
		\includegraphics[width=\columnwidth, angle=0,scale=1.0]{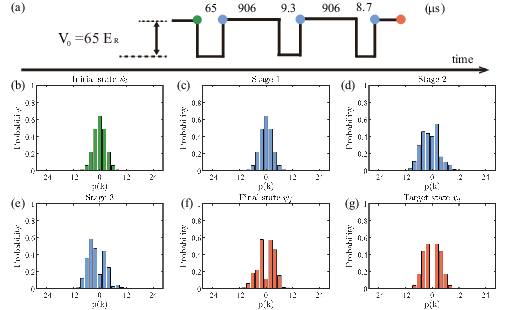}
		\caption{
			Time Evolution of the $q=0$ quasi-momentum state at a lattice depth of $65 E_R$, and the fidelity is about $91.33\%$. From left to right, and top to bottom, this corresponds to the time points marked with the same color in the above experimental sequence. It can be observed that the final result is already close to the Bloch states on the $p$-band.
		}    \label{p:state}
	\end{center}
\end{figure}

\begin{figure}[htbp]
	\begin{center}
		\includegraphics[width=\columnwidth, angle=0,scale=1.0]{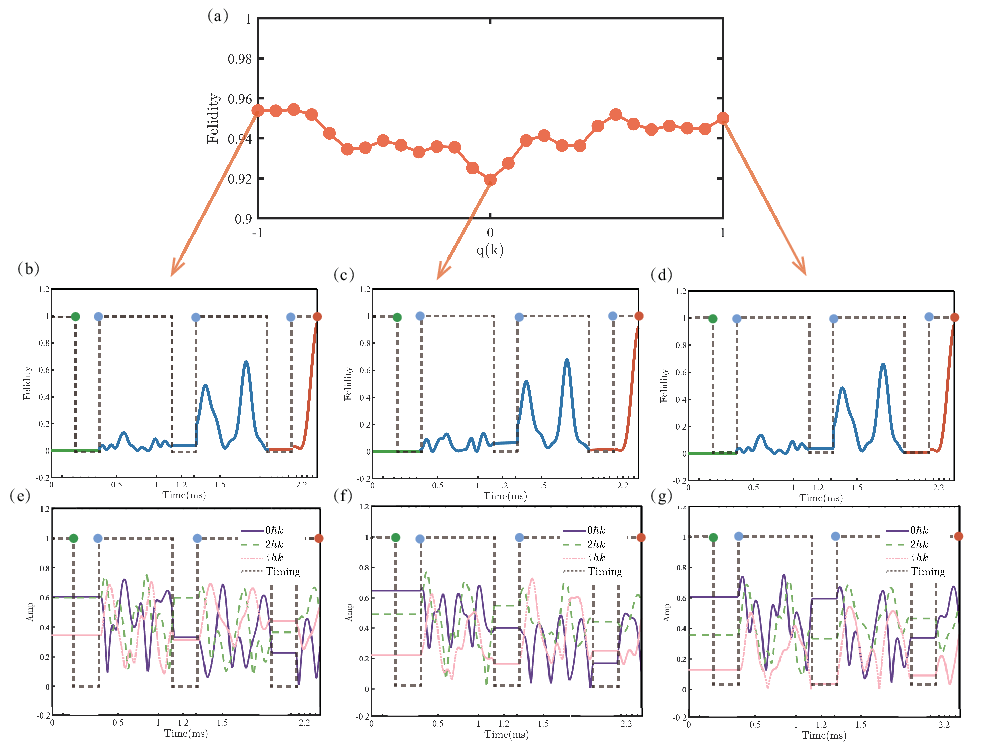}
		\caption{
			Evolution of fidelity and momentum components for three quasi-momentum states. (a) Fidelity of different quasi-momentum states, with final optimized parameters derived as the average across these states. (b-g) Time evolution of fidelity and momentum component amplitudes for the three states, with top and bottom subfigures depicting the same state; red arrows in (a) point to the corresponding states in (b-d).
		}    \label{p:f-q}
	\end{center}
\end{figure}

The high fidelity observed in Fig.~\ref{p:result} stems from the robust fidelity of each quasi-momentum state. To illustrate this, we present an example using optimized parameters for the $q=0$ state at a lattice depth of $65 E_R$, effectively demonstrating the loading process and its precision in reaching the target state. Fig.\ref{p:state}(b-g) illustrates this state’s time evolution, plotting the squared modulus of coefficients in the plane-wave basis across various stages, culminating in Fig.\ref{p:state}(f) for the final state and Fig.\ref{p:state}(g) for the target state, their comparison highlighting the evolution’s precision. Subplot colors denote distinct stages, aligned with the time sequence in Fig.\ref{p:state}(a). This visualization intuitively reveals how phase modulation and lattice switching shape the state’s composition. Over time, the initial state aligns closely with the target, achieving an average fidelity of $93.9\%$ and a $q=0$ state fidelity of $91.33\%$, as shown in Fig.~\ref{p:f-q}(a).

Fig.\ref{p:f-q} tracks the time evolution of three quasi-momentum states ($q=0, \pm1$), with Fig.\ref{p:f-q}(b-d) illustrating their fidelity trends and Fig.\ref{p:f-q}(e-g) detailing the corresponding momentum components. Fig.\ref{p:f-q}(a) indicates an average fidelity of $93\%$ across these states, highlighting consistent improvement driven by phase modulation. In Fig.\ref{p:f-q}(b-d), fidelity surges notably after each lattice-switching event, despite oscillations during lattice-on periods; these follow an upward trajectory, peaking in the final lattice-on segment. A black dashed line marks the time sequence, with interval proportions adjusted for clarity.

Fig.~\ref{p:f-q}(e-g) show the corresponding momentum component amplitudes, which exhibit oscillatory behavior during the lattice-on segments and remain unchanged during the lattice-off segments. This steady behavior during the off segments aligns with the fact that plane-wave states are eigenstates of the system in the absence of lattice potential, resulting in no evolution during these intervals. The three chosen quasi-momentum states occupy the dominant components of the final state, providing a clear demonstration of how phase modulation affects the state evolution. Together, these results emphasize the effectiveness of phase modulation in guiding the initial states toward the target state, significantly enhancing fidelity and ensuring a closer match with the desired high-orbital population.

\section{Discussion}
\label{sec:discussion}

\subsection{Fidelity depends on the number of quasi-momentum states}

\begin{figure}[htbp]
	\begin{center}
		\includegraphics[width=\columnwidth, angle=0]{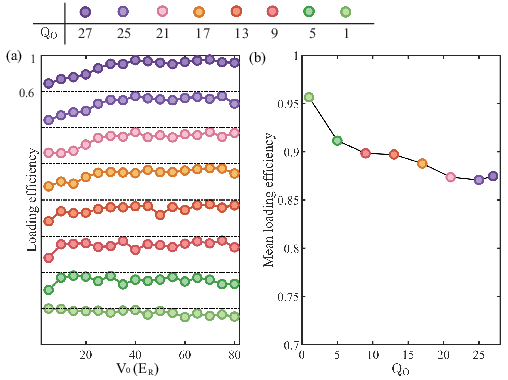}
		\caption{The relationship between loading efficiency and the number of quasi-momentum states $Q_\mathrm{O}$. (a) Illustrates the correlation between loading efficiency and lattice depth across various quasi-momentum occupancies. (b) Plots average loading efficiency against $Q_\mathrm{O}$, revealing an inverse dependence on the number of occupied states, aligning with expected physical behavior.
		}    \label{p:f-qn}
	\end{center}
\end{figure}

With the multi-phase optimization, we achieve a maximum loading efficiency of approximately $95\%$, which is still a little bit lower than the theoretical result of the BEC case~\cite{Zhou2018}. This is likely due to the occupation of multiple quasi-momentum states for Fermi gas. To investigate this, we vary the number of occupied quasi-momentum states, $Q_\mathrm{O}$. When $Q_\mathrm{O} = 1$, the system occupies only the $q = 0$ state, similar to the BEC case. As shown in Fig.~\ref{p:f-qn}(a), for higher $Q_\mathrm{O}$, the efficiency initially decreases, then increases with lattice depth, suggesting that deeper lattices may improve efficiency for Fermi gases.

Fig.~\ref{p:f-qn}(b) plots the average loading efficiency across all lattice depths, $F_\mathrm{m}$, as a function of $Q_\mathrm{O}$. At $Q_\mathrm{O} = 1$, $F_\mathrm{m}$ peaks at $95\%$, but declines with increasing $Q_\mathrm{O}$, suggesting that occupying multiple quasi-momentum states hampers efficiency. Beyond $Q_\mathrm{O} = 10$, $F_\mathrm{m}$ drops below $90\%$, reflecting the reduced loading efficiency of a Fermi gas compared to a BEC, attributable to its broader momentum distribution.

\subsection{Comparison with other loading schemes}

To better illustrate the advantages of our approach, we compare our results with those from other research groups. We compare our results with those reported in Zhou's work~\cite{Zhou2018}. Excluding the $a_2$ sequence in his study, our findings align well with other three cases, confirming the reliability of our simulation program.

We also compare our method with those used by other Fermi gas research groups. For three-dimensional optical lattices, Ref.~\cite{Venu2022} achieved an $85\%$ loading efficiency into the $p$-band using Raman transitions, completing the process in approximately $20\mu \mathrm{s}$.  While their approach offers faster loading, it requires additional laser systems and results in lower efficiency compared to our approach.

For two-dimensional optical lattices, Ref.~\cite{Jackson2023} used chemical potential adjustments to load atoms into high bands, achieving $15\%$ efficiency in $300 \mathrm{ms}$, which is slower and less efficient than our method. In a similar context, Ref.~\cite{Hachmann2021} achieved $60\%$ loading efficiency with population exchange and a superlattice, but this approach is more complex and less efficient than ours.

In a recent study, Ref.~\cite{Dale2024} extended their method to one-dimensional lattices, achieving $35\%$ loading efficiency into the $p$-band. While the loading time was not specified, it is likely to be on the order of hundreds of milliseconds, further demonstrating the higher efficiency of our method.

\begin{figure}[htbp]
	\begin{center}
		\includegraphics[width=\columnwidth, angle=0]{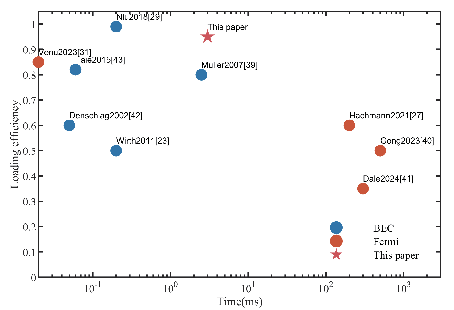}
		\caption{
			We have summarized the work of several groups, the red pentagram represents this work, and it is evident that our approach achieves the highest loading efficiency within the Fermi system, with relatively short loading times. In contrast, while the BEC system achieves even higher loading efficiency, it also benefits from shorter loading times.
		}    \label{p:exp compare}
	\end{center}
\end{figure}

We summarize the comparison results in Fig.~\ref{p:exp compare}, including the BEC case. Our loading method consistently outperforms alternatives in terms of efficiency, with our method also excelling in loading time, second only to the Raman transition technique. This is particularly advantageous for experiments studying the lifetime of atoms in high-orbital states.

Our approach is especially effective for one-dimensional optical lattices and can be adapted to other lattice configurations with minimal modifications. Unlike existing methods, which struggle with broad momentum distributions, our method efficiently loads Bloch states even with large momentum spreads, offering a more comprehensive solution for creating quantum states across a range of systems.

\section{Summary}
\label{sec:summary}

In this work, we developed a one-dimensional optical lattice $p$-band loading technique for ultracold Fermi gases. By overcoming the broad momentum distribution of Fermi gases, we used a \texttt{MATLAB} global optimization package to determine the optimal time sequence and phase change.

Initially, we adapted the BEC loading method~\cite{Liu2011,Niu2018,Jin2022}, but found it less effective for Fermi systems. By introducing phase as an optimization parameter, we significantly improved loading efficiency, which increased with lattice depth. We also explored the impact of the number of occupied quasi-momentum states on loading efficiency, confirming that fewer occupied states lead to higher efficiency, resembling the BEC scenario.

Our theoretical analysis shows that lattice phase influences atom loading, even with broad momentum distributions, due to the distinct parities of Bloch states. These findings contribute to understanding loading techniques for ultracold Fermi gases in optical lattices. Building on this theory, our group is now pursuing experiments on high-band relaxation and few-body physics, providing a foundation for future studies in many-body physics.

\section*{ACKNOWLEDGMENTS}

This work is supported by  the National Key Research and Development Program of China under Grant No.2022YFC2204402, the National Natural Science Foundation of China under Grant No.12174458, Guangdong Provincial Quantum Science Strategic Initiative under Grant No.GDZX2203001, No.GDZX2303003, Shenzhen Science and Technology Program under Grant No.JCYJ20220818102003006, the Fundamental Research Funds for the Central Universities, Sun Yat-sen University (24xkjc015).
\appendix
\section{PLANE-WAVE DECOMPOSITION METHOD}
\label{app:method}

In our setup, the one-dimensional optical lattice is created using two laser beams that form an interference pattern with a specific angle between them, resulting in a simplified optical lattice potential:
\begin{equation}\label{eq:1}
	V(x)=-s E_R \cos ^2\left(\frac{\pi x}{d}\right)=\sum_G V_G e^{i G x},
\end{equation}
where, $E_R=\hbar^2(2 \pi \sin \phi)^2 /\left(2 m\lambda^2\right)$ is the recoil energy, $\phi$ is angular separation of two laser beams, $\lambda$ is the laser wavelength, and $d$ is the lattice constant. The latter part of this expression can be understood as decomposing the periodic structure into a series of superimposed plane waves, $G = 2l\pi /d$ is reciprocal lattice vectors, $l$ takes integer values, $V_G$ is the intensity coefficient for different plane wave.

Although Bloch states originate from a single-particle model, we model non-interacting Fermi gases as independent single particles. Thus, the Hamiltonian can be expressed as follows, with $\hbar = 1$:
\begin{equation}
	\left(\frac{1}{2 m} p^2+\sum_G V_G e^{i G x}\right) \psi(x)
	=\varepsilon \psi(x).
	\label{eq:2}
\end{equation}

The solutions to this Schrödinger equation are known as Bloch states, which can also be expressed as a superposition of plane waves:
\begin{equation}
	\begin{aligned}
		&\psi(x) = |n, q\rangle=u_{n, q}(x) e^{i q \cdot x}\\ &
		= \sum_G C_{q+G}^n \frac{e^{i(q+G) x}}{\sqrt{N_s d}} = \sum_l c_{n, q, l}|2 l k+q\rangle,
	\end{aligned}
	\label{eq:3}
\end{equation}
where, $n$ is the band index, $|2 l k+q\rangle$ presents plane waves basis for different reciprocal lattice vectors, $k = \pi / d$ is the wave number of lattice light, $N_s$ is the number of lattice sites, $q$ is the quasi-momentum, restricted to the values of: $q=2 m \pi /\left(N_s d\right)$, $m$ takes integer values. Each plane wave has a wave vector corresponding to the quasi-momentum and reciprocal lattice vector, i.e., $p = 2 l \pi /d+q$, subsequently, we describe the entire physical process in this basis.

\section{EXPERIMENT SETUP}
\label{app:experiment setup}

Our experimental scheme aims to investigate high-orbital loading in an optical lattice, focusing on precise control of the lattice phase to alter the parity of Bloch states. The current experimental setup consists of ultracold lithium-6 atoms~\cite{Peng2024}, with the optical lattice constructed using lasers with a wavelength $ \lambda = 1064 ~ \mathrm{nm}$. The two laser beams forming the optical lattice intersect at an angle of $2 \theta=20^{\circ}$, resulting in a lattice constant $d=\lambda /(2 \sin \theta)=3.06 ~ \mathrm{\mu m}$ and the recoil energy $E_R=(2 \pi \hbar \sin \theta)^2 / 2 m \lambda^2 \approx 883 ~ \mathrm{Hz}$, the maximum lattice depth value is $127.3 E_R$. The frequency ratios in the three directions are $\omega_x: \omega_y: \omega_z \approx 5.76: 1: 514$.

\begin{figure}[htbp]
	\begin{center}
		\includegraphics[width=\columnwidth, angle=0, scale=1.0]{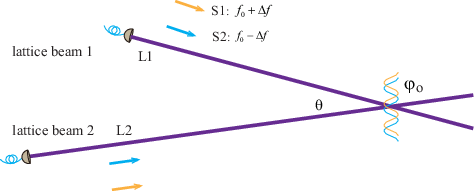}
		\caption{The modified optical setup, creating a fixed phase difference to meet the requirements for phase adjustment.
		}    \label{p:exp}
	\end{center}
\end{figure}

To modulate the optical lattice phase, we introduce an optical path difference between the two laser beams by horizontally displacing them along the lattice axis. The phase shift is given by the equations:
\begin{equation}
	\begin{gathered}
		\varphi_1=k_1\left(L_1-L_2\right), \\
		\varphi_2=k_2\left(L_1-L_2\right),
	\end{gathered}
	\label{eq:30}
\end{equation}
where $k_1$ and $k_2$ are the wave vectors, and $L_1$ and $L_2$ are the optical paths of the two lattice beams. The phase of the lattice potential can be optimized by adjusting the laser frequency using an acousto-optic modulator~\cite{Zhang2019}, as shown in Fig.~\ref{p:exp}.

\section{CASE OF FIXED PHASE CHANGE TO $3 \pi /4$}
\label{app:3pi/4}

\begin{figure}[htbp]
	\begin{center}
		\includegraphics[width=\columnwidth, angle=0,scale=1.0]{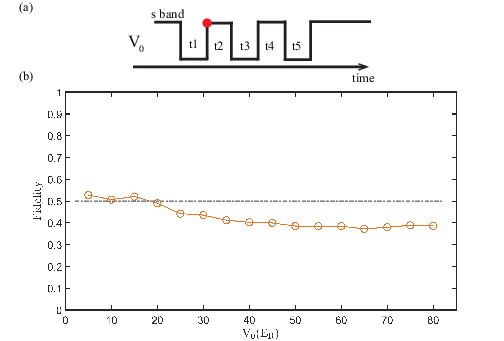}
		\caption{
			The optimization results for the fixed phase of $3\pi/4$ are shown. (a) Time sequence with phase transitions marked by red dots. (b) Optimized fidelity plotted against lattice depth.
		}    \label{p:3pi/4}
	\end{center}
\end{figure}

Based on the theoretical model, we modulated the optical lattice phase to $3\pi/4$ and performed the simulation. As shown by the yellow line in Fig.~\ref{p:3pi/4}(b), the maximum loading efficiency reaches $53\%$ at shallow lattice depths and gradually decreases, stabilizing around $40\%$. This contrasts with the first scenario, indicating that phase modulation significantly impacts loading efficiency across the full range of lattice depths.

Comparing this with the first case, fixing the phase at $3 \pi /4$ enhances loading efficiency at lower lattice depths but reduces it at higher depths, with both scenarios intersecting at $V_0 \approx 37E_R$. This suggests that different loading strategies depend on lattice depth, and phase alterations play a crucial role. However, simply adjusting a fixed phase does not yield optimal results, highlighting the need for phase optimization.

%

\end{document}